\documentclass{article}

\usepackage{spconf,amsmath,amssymb,graphicx}
\usepackage{booktabs}
\usepackage{hyperref}
\usepackage{float}
\usepackage[skip=1pt]{caption}

\usepackage{color}

\title{Learning domain-invariant classifiers for infant cry sounds}
\name{ Charles C. Onu\textsuperscript{1,2,3*}, Hemanth K. Sheetha\textsuperscript{1,2*}, Arsenii Gorin\textsuperscript{1}, Doina Precup\textsuperscript{2,3}}
\address{\textsuperscript{1}Ubenwa Health\\ \textsuperscript{2}Mila-Quebec Artificial Intelligence Institute\\  \textsuperscript{3}McGill University}

\begin{document}
%
\maketitle
\begin{abstract}

The issue of domain shift remains a problematic phenomenon in most real-world datasets and clinical audio is no exception. In this work, we study the nature of domain shift in a clinical database of infant cry sounds acquired across different geographies. We find that though the pitches of infant cries are similarly distributed regardless of the place of birth, other characteristics introduce peculiar biases into the data. We explore methodologies for mitigating the impact of domain shift in a model for identifying neurological injury from cry sounds. We adapt unsupervised domain adaptation methods from computer vision which learn an audio representation that is domain-invariant to hospitals and is task discriminative. We also propose a new approach, target noise injection (TNI), for unsupervised domain adaptation which requires neither labels nor training data from the target domain. Our best-performing model significantly improves target accuracy by 7.2\%, without negatively affecting the source domain.
\end{abstract}
%
\section{Introduction}
\label{sec:intro} 
\def\thefootnote{*}\footnotetext{Equal contribution}

Training a neural net for a new task can be expensive. Models typically contain hundreds of millions of parameters requiring immense compute but also large amounts of labelled data which can be costly to obtain in a clinical setting. When solving a given task, we ideally want to build a model that performs at similar accuracy when deployed in new settings or domains. In reality, it is rarely the case that distributions of source and target data are the same \cite{candela2009dataset}, typically resulting in inconsistent model performance \cite{torralba2011unbiased, yosinski2014transferable}. This bias in a dataset can be due to many factors including variations in sensors used to capture data, environmental noise, acquisition protocols, and many more. 


Domain adaptation (DA) aims to address the impact of dataset bias on generalization. In pursuing DA, we want to learn a cross-domain classifier that performs well in both the source and the target domain by mitigating the distributional shift. The core idea in most DA algorithms is to simultaneously or adversarially solve a classification task while learning a domain-invariant representation. This is typically achieved by minimizing a loss consisting of terms for the classification error as well as a measure of the statistical difference between the 2 distributions. Divergence measures used in the latter include mean maximum discrepancy (MMD) \cite{tzeng2014deep},  maximum mean feature norm discrepancy \cite{xu2019larger}, and correlation distance \cite{sun2016deep, sun2016return}. The H-divergence has also been employed in an adversarial objective by using a gradient reversal on a domain classifier \cite{ganin2016domain} or adversarial discriminative domain adaptation \cite{tzeng2017adversarial}. The Wasserstein distance has also proved a useful divergence in adversarial domain adaptation \cite{shen2018wasserstein}. 

\begin{figure}[b!]
\begin{minipage}[b]{0.45\textwidth}
  \centering
  \centerline{\includegraphics[width=\linewidth]{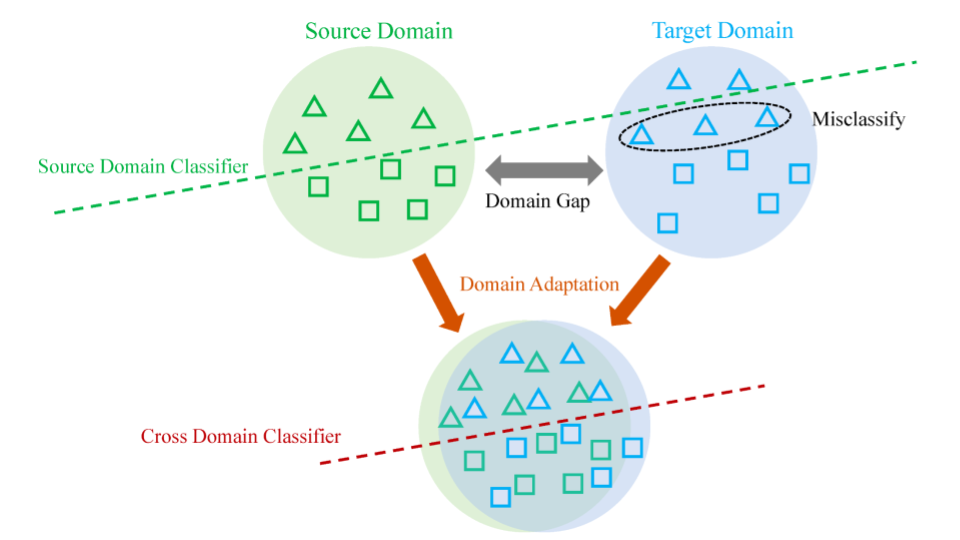}}
\end{minipage}
\caption{In domain adaptation, we aim to build a cross-domain classifier that generalizes in a consistent fashion regardless of biases in the data samples. Image source \cite{shi2022deep}.}
\label{da-illustration}
\end{figure}

\begin{figure*}[ht!]
\label{fig:domain-shift-id}
\begin{minipage}[b]{0.33\textwidth}
  \centering
  \centerline{\includegraphics[width=\linewidth]{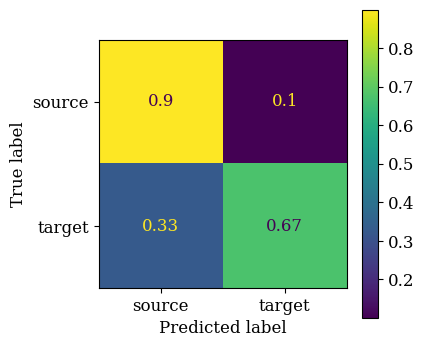}}
  \centerline{(a)}\medskip
  \label{fig:1a}
\end{minipage}
\begin{minipage}[b]{0.33\textwidth}
  \centering
  \centerline{\includegraphics[width=\linewidth]{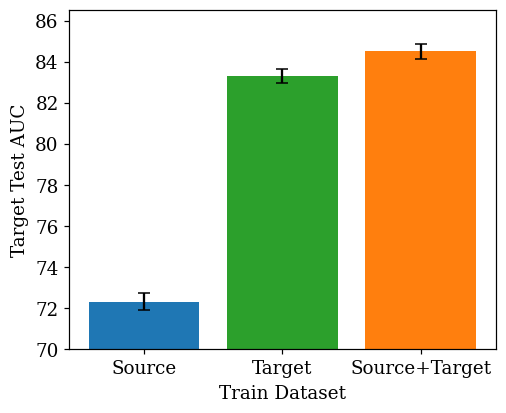}}
  \centerline{(b)}\medskip
  \label{fig:1b}
\end{minipage}
\hfill
\begin{minipage}[b]{0.33\textwidth}
  \centering
  \centerline{\includegraphics[width=\linewidth]{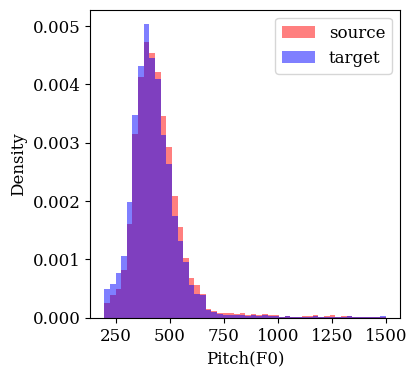}}
  \centerline{(c)}\medskip
  \label{fig:1c}
\end{minipage}
\caption{Identifying and understanding the nature of domain shift between source and target hospital datasets of infant crying. (a) Confusion matrix of the domain classifier indicating that model can correctly guess where a sample came from (b) Cross-hospital generalization showing that model train on source data only doesn't generalize well to target domain. (c) The distribution of cry pitch in the source and target domain suggests that dataset bias does not emanate from the cry signals.}

\label{fig:identification}
\end{figure*}
Here, we are interested in domain adaptation in the context of identifying signs of neurological injury from audio recordings of infant cries. Over a span of 3 years, the Ubenwa clinical study \cite{gorin2023selfsupervised} collected cry recordings across hospitals in 3 countries (Brazil, Canada, and Nigeria) for this problem. Multiple prior work have developed neuro injury detection models from cry sounds using neural transfer learning\cite{onu2020neural} and self-supervised learning\cite{gorin2023selfsupervised}. Although these methods show effectiveness on in-domain test sets, they fail to generalize as well to new hospital data.

In this work, we identify and study patterns of domain shift using this international database of infant cry recordings and explore methods for domain adaptation. We show that DA methods from computer vision can be repurposed and applied to infant cry audio. By experimenting with 5 different methods we illustrate that the best methods not only improve target accuracy but also accuracy in the source domain. Secondly, we validate previous clinical findings about the newborn cry as a universal language -- the pitch of baby cries is similarly distributed regardless of geography. We propose a relatively simple and promising approach for DA in infant cry data. Our method requires no architectural changes nor complex, min-max optimization, employs a simple cross-entropy loss function, and requires neither labels nor cry recordings from the target domain -- only target noise samples \cite{NANNI2020101084}. Experiments show that this is a promising direction worth exploring further in future work.


\begin{table*}[htbp]
\centering
\resizebox{\textwidth}{!}{
\begin{tabular}{cccccccc}
    \hline
    Methods & \begin{tabular}[c]{@{}c@{}}Source \\ test AUC\end{tabular} & \begin{tabular}[c]{@{}c@{}}Target \\ test AUC\end{tabular} & Loss Function & \begin{tabular}[c]{@{}c@{}}Requires architecture \\ modification\end{tabular} & \begin{tabular}[c]{@{}c@{}}Requires cry samples \\ from target domain\end{tabular} & \begin{tabular}[c]{@{}c@{}}Target test \\ AUC improvement\end{tabular} & \begin{tabular}[c]{@{}c@{}}Source test \\ AUC improvement\end{tabular} \\ \hline
    No DA & 80.00 $\pm$ 1.9 & 72.30 $\pm$ 0.9 & - & - & - & - & - \\ \hline 
    EM & 76.00 $\pm$ 3.3 & 72.86 $\pm$ 1.8 & CE + EM & no & yes & 0.7\% & -5.3\% \\
    Unsupervised BN & 77.16 $\pm$ 1.9 & 73.04 $\pm$ 0.9 & CE & no & yes & 1.00\% & -3.55\% \\
    SAFN & 86.67 $\pm$ 1.1 & 75.85 $\pm$ 1.3 & CE + DC & yes & yes & 4.89\% & \textbf{8.33\%} \\
    HAFN & 85.47 $\pm$ 1.9 & 73.98 $\pm$ 1.9 & CE + DC & yes & yes & 2.30\% & 6.80\% \\
    SymNets & 83.54 $\pm$ 0.7 & 77.52 $\pm$ 2.1 & CE + DC + DD & yes & yes & \textbf{7.20\%} & 4.42\% \\
    TNI & 84.09 $\pm$ 0.2 & 72.94 $\pm$ 2.0 & CE & no & no (only noise) & 0.87\% & 5.1\% \\ 
    \hline
\end{tabular}
}
\caption{Performance of different unsupervised domain adaptation methods. CE=Cross Entropy, EM=Entropy Minimization, DC=Domain Confusion, DD=Domain Discrimination. Other abbreviations are as used in this paper.}
\label{tab:main-result}

\end{table*}

\section{Domain Shift Identification}
\label{sec:domain_shift_id}

In this section, we present our approach to studying the presence and patterns of domain shift in infant cry audio

\subsection{Name-the-hospital challenge}
We first ask the question: Can we build a model to identify which hospital or location a given cry recording came from? Domain shift is characterized by bias in the datasets. If such bias exists, we can build a classifier to identify which hospital a sample comes from \cite{5995347}. If it doesn't, the classifier should struggle to distinguish recordings from different hospitals. We thus train a domain classifier using the hospital as target labels. Our model was a simple convolutional neural net and the result showed that it could accurately identify the originating hospital of a cry recording (Fig \ref{fig:identification}a).

\subsection{Cross-hospital generalization}
We further conducted experiments on cross-hospital generalization. Here we train a classifier (CNN14 \cite{kong2020panns}) on the source hospital data and test it on the target. If there is no domain shift between hospitals, the model should generalize to the target domain. However, results (Fig \ref{fig:identification}b) show that the model trained on only source data performs worse than models training using (in-domain) target training data -- another sign of domain shift.

\subsection{Distribution of fundamental frequency}
\label{sec:pitch-distribution}
Thirdly, to better understand the nature of domain shift, we compare the distribution of features of data from the two domains. In the case of an infant cry, one of the most important features is pitch or fundamental frequency \cite{corwin1996infant}. It corresponds to the rate of vibration of the vocal cords during a cry expiration and defines the harmonic properties of infant cries including the formants \cite{golub1985physioacoustic}. We compare pitch distributions across the two domains as a way of isolating where bias in the dataset might be coming from. To do this, we use a cry activity detection model to segment cry sounds, then run a pitch estimator (CREPE \cite{8461329}) to obtain the pitch tracks per cry utterance. As seen in the overlapping histograms in figure \ref{fig:identification}c, pitch is similarly distributed across cry samples from the different hospitals, suggesting that the bias is coming from other factors such as background noises or acoustic properties of the recording environment. This finding is consistent with previous research showing that the newborn cry is a pure signal not altered by a baby's genes or birthplace.

\section{Learning Domain-Invariant Classifiers}

\subsection{Encoder \& Classifier Architecture}
Our model consists of a backbone encoder followed by a classifier. We adopt, as an encoder, a CNN14 \cite{kong2020panns} pre-trained on VGGSound\cite{chen2020vggsound}, a large generic audio dataset. We chose VGGSound for pre-training as it consists of sounds rather than speech which is more transferable to cry sounds. For the classifier, we add one feed-forward layer on top of the backbone encoder. Unless otherwise specified, we only tune the batch norm layers of the backbone encoder(the rest of the encoder layers are frozen) and also train the classifier from scratch as it was found to be efficient in low resource transfer learning\cite{gorin2023selfsupervised} and confirmed in our preliminary experiments.

\begin{figure*}[h]
\begin{minipage}[b]{0.5\textwidth}
  \centering
  \centerline{\includegraphics[width=\linewidth]{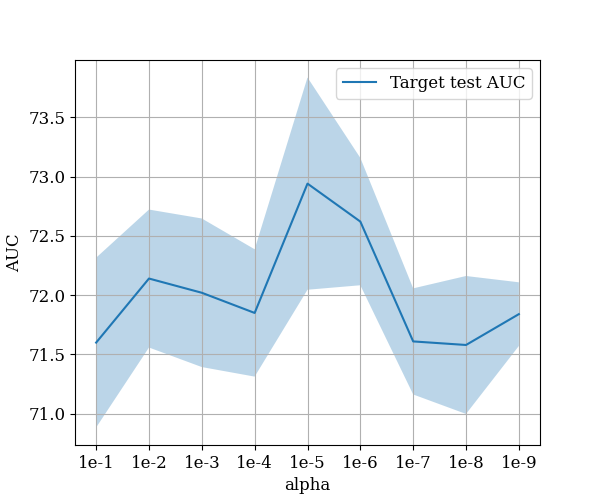}}
    \centerline{(a)}\medskip
  \label{fig:3a}
\end{minipage}
\begin{minipage}[b]{0.5\textwidth}
  \centering
  \centerline{\includegraphics[width=\linewidth]{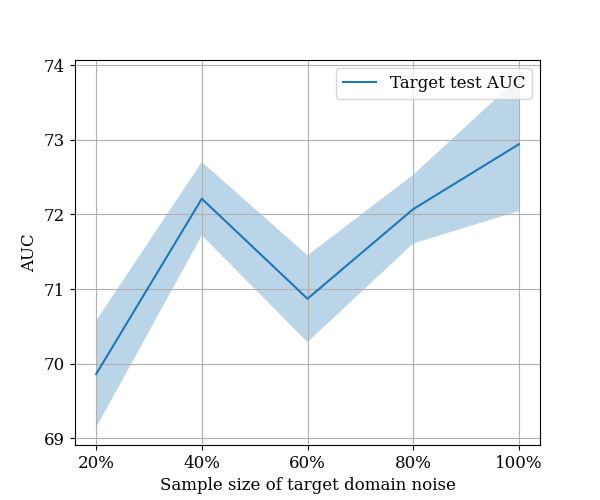}}
  \centerline{(b)}\medskip
  \label{fig:3b}
\end{minipage}
\caption{(a) AUC in target domain for different values of $\alpha$, the weight on the target domain noise (b) AUC in target domain as the amount of target domain noise is increased}
\label{fig:noise_sample_size}
\end{figure*}

\subsection{Unsupervised Domain Adaptation Methods}

Here we introduce the different domain adaptation methods tailored to infant cry audio.\\
\\
\noindent \textbf{Unsupervised Batch Normalization(BN)}\cite{li2016revisiting} assumes a model that is pre-trained on source domain data. We freeze the whole pre-trained model and classifier (including affine params of the batch normalization layers). Fine-tuning consists of updating the running mean and variance of the corresponding batch normalization layers on unlabeled target domain data. \\


\noindent \textbf{Entropy Minimization(EM)}\cite{NIPS2004_96f2b50b}\cite{zhang2018importance} is a semi-supervised learning method that simultaneously trains the model using labeled data and unlabeled data together. We optimize the following extra term along with the cross-entropy loss which tries to increase the confidence of the classifier output of target examples by minimizing the entropy:
\begin{equation}
\min_{G} \min_{C} \mathbb{E}_{\textbf{x}\sim p_t(x)}C(G(\textbf{x}))
\end{equation}
%
%

\noindent where G is the backbone encoder,  C is the Classifier, and $p_t(x)$ is a distribution sampling from the target domain.\\ 

\noindent \textbf{Adaptive Feature Norm(AFN)}\cite{xu2019larger} introduces the Maximum Mean Feature Norm Discrepancy(MMFND) metric which measures the gap between the two domain distributions. This measure helps in reducing the domain shift. The optimization objective consists of the cross-entropy loss of labeled source examples and has another loss that reduces the statistical domain gap between source and target domains using the AFN objective. In hard AFN (HAFN), a restrictive scalar R restricts the mean feature norms of our domains, while in stepwise AFN (SAFN), a step size of $\Delta r$ is introduced to encourage larger norms that are more informative and leads to better target performance.\\

\noindent \textbf{Domain Symmetric Networks(SymNets)}\cite{zhang2019domainsymmetric} uses adversarial training to make domains invariant and tasks discriminative. While other methods only use a source domain classifier, SymNets makes use of a symmetric design that includes an explicit task classifier for the target domain in addition to the source domain classifier. The target classifier is trained to improve target predictions using source labels, making them more task discriminative. Their optimization objective comprises cross-domain training terms, two-level domain confusion losses for domain invariance, and domain discrimination and entropy minimization. 

\subsection{Target noise injection for domain adaptation}
Sequel to the insights from section \ref{sec:pitch-distribution}, we propose a relatively simple approach for domain adaptation -- target noise injection (TNI). In this method, we segment, extract, and inject target domain noise into source samples during training. The intuition is that if the environmental noise in the target hospital is the primary source of domain shift, then training the source classifier to be robust to such noise will enable the classifier to learn effective, cross-domain representations. This approach has practical benefits. Data collection only requires recording noises in the target environment -- no need for labels or actual cry recordings -- which can be much cheaper and faster to accomplish. During model training, this method requires no new loss function (cross-entropy is sufficient), no change to model architecture, and no complicated training paradigm like min-max optimization. Each audio sample, $s$ in a batch is augmented as $s' = s + \alpha n$, where n is a sample of target domain noise, and $\alpha \in [0, 1]$ is a hyperparameter.

\section{Experiments}

\subsection{Experimental setup and training details}
This study uses a subset of the Ubenwa newborn cry database collected from hospitals in Brazil, Canada, and Nigeria. Each cry recording is annotated as either healthy or neurological injury based on clinical exams conducted by doctors. We select a source hospital and a target hospital for the purposes of this work. The data is split into train, validation, and test sets making sure that all recordings from an individual patient belong to only one set. There were a total of 406 patients in the source hospital and 1,507 in the target hospital. We report the mean and standard error of AUC across 5 training runs. All models were trained using the Adam optimizer with a batch size of 32. For each method, learning rates for the backbone encoder and classifier were tuned as separate hyperparameters using the validation sets. For HAFN, R is set to 30, and for SAFN, $\Delta r$ is set to 0.2.




\subsection{Results}
Our results for unsupervised domain adaptation (UDA) on infant cry data for classifying neurological injury are summarized in Table \ref{tab:main-result}. We see that all domain-adapted models outperform the unadapted model in the target domain. This is consistent with previous work in computer vision where these methods were found to to be effective. The biggest improvement in target AUC of 7.2\% was achieved by SymNet, while the smallest improvements were observed with entropy minimization (0.7\%) and target noise injection (0.87\%). Though TNI achieves, a relatively small improvement over the unadapted model, sample size experiments (Fig \ref{fig:noise_sample_size}b) indicate that the model is far from saturated and it improves as more target noise is collected. In Fig \ref{fig:noise_sample_size}a, we see as well that the value of $\alpha$ impacts the quality of the adaptation. Too small values would mean not enough signal to reap benefits, while too large values could degrade performance due to too much noise.

When adapting models in the clinical setting, we are interested in not only improving model performance in the target hospital but also preserving source hospital performance, such that the model remains useful across the board. We find that only 2 methods negatively impact source AUC -- entropy minimization and unsupervised BN. We suspect that, in the case of BN, this is due to its post-hoc nature i.e., the adaptation step is applied after the model has been trained as opposed to simultaneous training and adaptation.




\section{Conclusion}
In this paper, we demonstrated the effectiveness of unsupervised domain adaptation methods in neuro-injury cry models. These methods were originally designed for computer vision tasks, but we show that they can be accurately repurposed for clinical audio like infant cry sounds. Through insights generated in this study, we propose a new unsupervised domain adaptation technique that requires neither training examples nor training labels in the target domain -- only noise recordings. It is practically convenient and shows promise worth exploring further in future work and with more data.



\bibliographystyle{IEEEbib}
\bibliography{main}

\begin{thebibliography}{10}

\bibitem{candela2009dataset}
J~Quinonero Candela, Masashi Sugiyama, Anton Schwaighofer, and Neil~D Lawrence,
\newblock ``Dataset shift in machine learning,''
\newblock {\em The MIT Press}, vol. 1, pp. 5, 2009.

\bibitem{torralba2011unbiased}
Antonio Torralba and Alexei~A Efros,
\newblock ``Unbiased look at dataset bias,''
\newblock in {\em CVPR 2011}. IEEE, 2011, pp. 1521--1528.

\bibitem{yosinski2014transferable}
Jason Yosinski, Jeff Clune, Yoshua Bengio, and Hod Lipson,
\newblock ``How transferable are features in deep neural networks?,''
\newblock {\em Advances in neural information processing systems}, vol. 27,
  2014.

\bibitem{tzeng2014deep}
Eric Tzeng, Judy Hoffman, Ning Zhang, Kate Saenko, and Trevor Darrell,
\newblock ``Deep domain confusion: Maximizing for domain invariance,''
\newblock {\em arXiv preprint arXiv:1412.3474}, 2014.

\bibitem{xu2019larger}
Ruijia Xu, Guanbin Li, Jihan Yang, and Liang Lin,
\newblock ``Larger norm more transferable: An adaptive feature norm approach
  for unsupervised domain adaptation,''
\newblock in {\em Proceedings of the IEEE/CVF international conference on
  computer vision}, 2019, pp. 1426--1435.

\bibitem{sun2016deep}
Baochen Sun and Kate Saenko,
\newblock ``Deep coral: Correlation alignment for deep domain adaptation,''
\newblock in {\em Computer Vision--ECCV 2016 Workshops: Amsterdam, The
  Netherlands, October 8-10 and 15-16, 2016, Proceedings, Part III 14}.
  Springer, 2016, pp. 443--450.

\bibitem{sun2016return}
Baochen Sun, Jiashi Feng, and Kate Saenko,
\newblock ``Return of frustratingly easy domain adaptation,''
\newblock in {\em Proceedings of the AAAI conference on artificial
  intelligence}, 2016, vol.~30.

\bibitem{ganin2016domain}
Yaroslav Ganin, Evgeniya Ustinova, Hana Ajakan, Pascal Germain, Hugo
  Larochelle, Fran{\c{c}}ois Laviolette, Mario Marchand, and Victor Lempitsky,
\newblock ``Domain-adversarial training of neural networks,''
\newblock {\em The journal of machine learning research}, vol. 17, no. 1, pp.
  2096--2030, 2016.

\bibitem{tzeng2017adversarial}
Eric Tzeng, Judy Hoffman, Kate Saenko, and Trevor Darrell,
\newblock ``Adversarial discriminative domain adaptation,''
\newblock in {\em Proceedings of the IEEE conference on computer vision and
  pattern recognition}, 2017, pp. 7167--7176.

\bibitem{shen2018wasserstein}
Jian Shen, Yanru Qu, Weinan Zhang, and Yong Yu,
\newblock ``Wasserstein distance guided representation learning for domain
  adaptation,''
\newblock in {\em Proceedings of the AAAI Conference on Artificial
  Intelligence}, 2018, vol.~32.

\bibitem{shi2022deep}
Yongjie Shi, Xianghua Ying, and Jinfa Yang,
\newblock ``Deep unsupervised domain adaptation with time series sensor data: A
  survey,''
\newblock {\em Sensors}, vol. 22, no. 15, pp. 5507, 2022.

\bibitem{gorin2023selfsupervised}
Arsenii Gorin, Cem Subakan, Sajjad Abdoli, Junhao Wang, Samantha Latremouille,
  and Charles Onu,
\newblock ``Self-supervised learning for infant cry analysis,'' 2023.

\bibitem{onu2020neural}
Charles~C. Onu, Jonathan Lebensold, William~L. Hamilton, and Doina Precup,
\newblock ``Neural transfer learning for cry-based diagnosis of perinatal
  asphyxia,'' 2020.

\bibitem{NANNI2020101084}
Loris Nanni, Gianluca Maguolo, and Michelangelo Paci,
\newblock ``Data augmentation approaches for improving animal audio
  classification,''
\newblock {\em Ecological Informatics}, vol. 57, pp. 101084, 2020.

\bibitem{5995347}
Antonio Torralba and Alexei~A. Efros,
\newblock ``Unbiased look at dataset bias,''
\newblock in {\em CVPR 2011}, 2011, pp. 1521--1528.

\bibitem{kong2020panns}
Qiuqiang Kong, Yin Cao, Turab Iqbal, Yuxuan Wang, Wenwu Wang, and Mark~D.
  Plumbley,
\newblock ``Panns: Large-scale pretrained audio neural networks for audio
  pattern recognition,'' 2020.

\bibitem{corwin1996infant}
Michael~J Corwin, Barry~M Lester, and Howard~L Golub,
\newblock ``The infant cry: what can it tell us?,''
\newblock {\em Current problems in Pediatrics}, vol. 26, no. 9, pp. 313--334,
  1996.

\bibitem{golub1985physioacoustic}
Howard~L Golub and Michael~J Corwin,
\newblock ``A physioacoustic model of the infant cry,''
\newblock in {\em Infant crying: Theoretical and research perspectives}, pp.
  59--82. Springer, 1985.

\bibitem{8461329}
Jong~Wook Kim, Justin Salamon, Peter Li, and Juan~Pablo Bello,
\newblock ``Crepe: A convolutional representation for pitch estimation,''
\newblock in {\em 2018 IEEE International Conference on Acoustics, Speech and
  Signal Processing (ICASSP)}, 2018, pp. 161--165.

\bibitem{chen2020vggsound}
Honglie Chen, Weidi Xie, Andrea Vedaldi, and Andrew Zisserman,
\newblock ``Vggsound: A large-scale audio-visual dataset,'' 2020.

\bibitem{li2016revisiting}
Yanghao Li, Naiyan Wang, Jianping Shi, Jiaying Liu, and Xiaodi Hou,
\newblock ``Revisiting batch normalization for practical domain adaptation,''
\newblock {\em arXiv preprint arXiv:1603.04779}, 2016.

\bibitem{NIPS2004_96f2b50b}
Yves Grandvalet and Yoshua Bengio,
\newblock ``Semi-supervised learning by entropy minimization,''
\newblock in {\em Advances in Neural Information Processing Systems}, L.~Saul,
  Y.~Weiss, and L.~Bottou, Eds. 2004, vol.~17, MIT Press.

\bibitem{zhang2018importance}
Jing Zhang, Zewei Ding, Wanqing Li, and Philip Ogunbona,
\newblock ``Importance weighted adversarial nets for partial domain
  adaptation,'' 2018.

\bibitem{zhang2019domainsymmetric}
Yabin Zhang, Hui Tang, Kui Jia, and Mingkui Tan,
\newblock ``Domain-symmetric networks for adversarial domain adaptation,''
  2019.

\end{thebibliography}

\end{document}